\documentclass[12pt]{article}
\pdfoutput=1
\usepackage{amstext,amssymb,amsfonts, graphicx,color}
\usepackage{color}
\usepackage{amsmath,amsfonts,epsfig,color,latexsym}
\usepackage{amssymb}
\usepackage[english, USenglish]{babel}
\usepackage{epsfig}
\usepackage{graphicx}
\numberwithin{equation}{section}

\advance\voffset by -1.5cm
\advance\hoffset by -1.25cm
\definecolor{darkgreen}{rgb}{0,0.6,0}
\def\thefootnote{\fnsymbol{footnote}}

\def\e{\,{\rm e}}

\def\ln{\,{\rm ln}}

\def\be{\begin{equation}}
\def\ee{\end{equation}}
\def\ba{\begin{eqnarray}}
\def\ea{\end{eqnarray}}

\def\nn{\nonumber}

\def\<{\langle}
\def\>{\rangle}

\begin{document}


\thispagestyle{empty}
\renewcommand{\thefootnote}{\fnsymbol{footnote}}

{\hfill \parbox{4cm}{
}}

\bigskip

\begin{center} \noindent \Large \bf
Comments on Correlation Functions of Large Spin Operators and Null Polygonal Wilson Loops 
\end{center}

\bigskip\bigskip\bigskip

\centerline{ \normalsize \bf Carlos A. Cardona${}^{a,\,b}$
  \footnote[1]{\noindent \tt cargicar@iafe.uba.ar}} 

\bigskip

\centerline{\it ${}^a$ Instituto de Astronom\'ia y F\'isica del Espacio (CONICET-UBA)}
\centerline{\it
C.C. 67 - Suc. 28, 1428 Buenos Aires, Argentina}
\vspace{0.3cm}
\centerline{\it ${}^b$ Physics Department, University of Buenos Aires, CONICET.}
\centerline{\it Ciudad Universitaria, 1428 Buenos Aires, Argentina}

\bigskip\bigskip

\bigskip\bigskip

\renewcommand{\thefootnote}{\arabic{footnote}}

\centerline{\bf \small Abstract}
\medskip

{\small \noindent We discuss the relation between correlation functions of twist-two large spin operators and expectation values of Wilson loops along light-like trajectories. After presenting some heuristic field theoretical arguments 
suggesting this relation, we compute the divergent part of the correlator 
 in the limit of large 't Hooft coupling and large spins, 
using a semi-classical worldsheet which asymptotically looks like a GKP 
rotating string. We show this diverges as expected from the 
expectation value of a null Wilson loop, namely, as $(\ln{\mu^{-2}})^ 2$, 
$\mu$ being  a cut-off of the theory.}

\vspace{0.5cm}



\setcounter{tocdepth}{2}
\tableofcontents

\newpage
\setcounter{equation}{0}
\section{Introduction}

In the last ten years much progress has been made concerning  single trace operators in ${\cal N}=4$ SYM theory at 
both weak and strong coupling in the context of the $AdS$/CFT correspondence \cite{Maldacena:1997re}. On one hand, 
the integrability of the theory in the planar limit \cite{Bena}-\cite{BeisStau} has led to  substantial advance,
 mainly in the computation of anomalous dimensions of
these operators. Indeed, it has guided us to a deeper understanding on the spectrum of conformal anomalous dimensions 
\cite{GromKazVier1}-\cite{ArutFrol2}. On the other hand, at the strong coupling limit of the theory, single-trace 
twist-two operators have been suggested to correspond to string states with angular momentum in the $AdS_5$ sector of 
 string theory on $AdS_5\times S^5$ \cite{Klebanov02}. Concretely,  twist two operators of the form\footnote{In this equation $\{\cdots\}$ means totally symmetrized and $V^ {\mu}$ is an arbitrary vector.}
\be\label{Twistwo}
{\cal O}_S ={\rm Tr}(\Phi\bigtriangledown_{\{ \mu_1}...\bigtriangledown_{\mu_S\}}\Phi)V^{\mu_1}...V^{\mu_S }\,,
\ee
with
spin $S$ and scaling dimension for large spin given by
\be\label{FoldedSpin}
 \Delta=S+2+\gamma_S\,,\quad\gamma_S=f(\lambda)\ln S\,,
\ee
are described at strong coupling by a macroscopic rotating string on $AdS_5$. At large $S$, the leading contribution to the anomalous dimension comes from the classical solution, whose energy satisfies the same scaling with the spin as the dimension (\ref{FoldedSpin}) \cite{Klebanov02, {Kruczenski:2004wg}, {Jevicki:2007aa}}. 
\vspace{0.2cm}

Other important objects which have been extensively studied during the last years are Wilson loops made from null lines. 
These have several interesting properties. To mention just a few, they are related to scattering amplitudes in gauge
 theories \cite{AldayHR}-\cite{DrummondAU}, particularly to
deep inelastic scattering amplitudes in QCD \cite{Korchemsky:1992xv}. But more relevant for the purposes of this note, 
they can be used also  to compute the anomalous dimensions of the operators we would like to consider, namely, those of the form (\ref{Twistwo}). This has been done at weak coupling 
in \cite{Korchemsky89, Korchemsky92} and at strong coupling  in \cite{Kruczenski:2002fb}. The anomalous dimension 
 determines the the two-point correlation function of the corresponding operator up to a coupling dependent normalization factor (which can also dependent on the spin), suggesting that there is a 
relation between correlation functions of operators ${\cal O_S}$ and expectation values of Wilson loops, in a similar way 
as the relation found between the latter and correlation functions of scalars in \cite{Alday:2010zy}. In this note we 
explore some possible connections between these two objects, both at strong coupling and weak coupling, which have been
suggested before in 
\cite{Alday:2010ku,Georgiou:2010an}.
\vspace{0.2cm}

At strong coupling, the expectation value of a Wilson loop is given by the area of the worldsheet which ends
 on it at the boundary of $AdS_5$ \cite{Maldacena:1998im}. Light-like polygonal Wilson loops have been 
extensively studied at strong coupling by Maldacena $et~ al.$ 
\cite{ Alday:2010ku, Alday:2009yn, Alday:2009dv, Alday:2010vh}.
Expectation values of null Wilson loops contain ultraviolet divergences at weak and strong coupling. At weak coupling,
 they are associated with the fast moving particles through the Wilson light-like lines as well as 
with the existence of cusps, and at strong coupling, these are due to the infinite area of the world-sheet 
ending on the 
sides of the Wilson loops. On the other hand, correlators of large spin operators ${\cal O}_S$ have
 two types of divergences. One type comes from the light-like distance limit between the operator and  
the other one
 comes from the large spin limit. As we will see, at strong coupling these two are somehow related. 

   
The calculation of the n-point function at strong coupling in the leading  semiclassical approximation was shown in~\cite{tsev,b1,bt1} to be intrinsically related to finding an appropriate classical solution of a string which ends on the boundary of $AdS$.
Let $ V_{\Delta_i}(Y(z_i)),\, i=1,...,n$ be the $n-$vertex operators inserted at points
$z_i$ on the worldsheet corresponding to a folded string and $Y$ denoting the collective target space coordinates.\footnote{We consider  planar $AdS$/CFT duality, 
i.e. tree-level  string theory.} For large string  tension, the n-point function should be dominated by its semiclassical limit i.e. by the action and vertex evaluated at its stationary point
\be\
\langle\prod_{i=1}^n  V_{\Delta_i}(z_i) \rangle \sim e^{-\frac{R^2}{2\pi\alpha'}A}\,\prod_{i=1}^n  V_{\Delta_i}(Y^*(z_i))\,,
\ee
where $A$ is the string action on $AdS_5 \times S^5$ in conformal gauge, which in the semiclassical limit is reduced to the area of the worldsheet, and $Y^*(z)$ denotes the classical string solution. The information provided by the vertex operators on the left hand side is implicitly contained in the boundary conditions for the classical solution associated to the area $A$, since in the semiclassical approximation they act as delta-sources setting the boundary conditions. We will argue below that the contribution near the insertions of the vertex operators should be finite in the large spin limit.
For the case of the two-point function, the appropriate classical solution corresponds to some analytical continuation of the GKP string \cite{Georgiou:2010an,tsev}, which reaches the boundary of $AdS$ when the spin of the string goes to infinity. This implies that, in the limit we are considering, the area of the worldsheet diverges and should be somehow regularized, in the same way as for expectation values of Wilson lines
 at strong coupling.
This should
also  happen in the classical solutions corresponding to higher point functions. Hence, the n-point functions split in a divergent contribution and a finite (regularized) part,
\be
\langle\prod_{i=1}^n  V_{\Delta_i}(z_i) \rangle \sim {\rm exp}\left[-\frac{R^2}{2\pi\alpha'}(A_{\text{div}}+A_{\text{reg}})+\sum_i^n\ln\{ V_{\Delta_i}(Y^*(z_i))\}\right]\,.
\label{1.4}
\ee
In this note we will study the divergent part of the correlator and we will show that it scales in the same way as the null polygonal Wilson loop associated with it\footnote{ In a recent paper \cite{Kazama:2011cp} the finite contribution $A_{\text{reg}}$ of the 3-point  correlator has been computed by exploiting the integrability of string theory on $AdS_3$ through the use of the Pohlmeyer reduced system.}. It is worth  mentioning here that our approach to compute the divergent factor of the correlation function is very much in the spirit of \cite{AldayHR}.

The presentation is organized as follows.
In section 2 we will give some hints from the field theory perspective supporting the relation between correlation functions of large spin operators (\ref{Twistwo}) and expectation values of null polygonal Wilson loops.
In section 3 we will summarize the Pohlmeyer reduction of strings on $AdS_3$. This is a useful technique which reduces
 the problem of solving the string equations of motion to that of finding a solution of a generalized Sinh-Gordon equation. 
This reduction is particularly powerful when we want to explore and use the integrability of  string theory on 
$AdS$ spaces.   In section 4 we will use the Pohlmeyer reduced fields in order to study worldsheet solutions which contain
 asymptotically folded strings at large spin and from these we will compute the  corresponding regularized area.
 
\section{Field Theory Considerations}

In this section we will review some evidence suggesting the relation between the large spin limit of correlators involving the fields (\ref{Twistwo}) and expectation values of null polygonal Wilson loops.
We are considering operators of the schematic form ${\cal O}_S = {\rm Tr}[\Phi (V^{\mu}\bigtriangledown_{\mu})^S \Phi ] $, which are characterized by the spin number $S$ and the conformal weight $\Delta$. In the large spin limit, we will compute the correlation functions in the region in which the emitted particles tend to follow light-like directions. In order to make direct contact with Wilson loops, we recall that the operator $O_S$ arises in a power series expansion of the following gauge-invariant bi-local operator
\be\label{wilsonxp}
W(V)={\rm Tr}
\left(\Phi(0)\e^{\int_0^{V}A_{\mu}dV^{\mu}}\Phi(V^{\mu})
\e^{-\int_0^{V}A_{\mu}dV^{\mu}}\right)=\sum_{S=0}^{\infty}\frac{1}{S!}{\cal O}_S(V^{\mu})\,.
\ee 

On one hand, we can computed the leading divergent factor of the following expectation value \cite{Georgi74},
\be\label{evo}
\langle p| {\cal O}_S|p\rangle\sim (p\cdot V)^S\left(\frac{\Lambda}{\mu}\right)^{\gamma_S}\,,
\ee 
where ${\Lambda}$ and ${\mu}$ are ultraviolet and infra-red cut-offs respectively.
(\ref{evo}) means we can compute the expectation value of $W_V$ as, 
\be\label{Wexpan}
\langle p|W_{V}|p\rangle=\sum_0^S\frac{1}{S!}(p\cdot V)^S\left(\frac{\Lambda}{\mu}\right)^{-f(\lambda)\ln S}=\sum_S\frac{1}{S!}(p\cdot V)^S S^{-f(\lambda)\ln(\Lambda/\mu)}\,.
\ee
By using the following identity,
\be
\sum_n\frac{a_n}{n!}(x)^n =x^{\alpha}\e^{x}\,,~~~~(x\to\infty)~~~\mbox{if}\quad a^n\sim n^{\alpha}\,,(n\to\infty)
\ee
we can conclude that in the large spin limit the series is dominated by,
\be\label{limitseries}
\langle p|W_{V}|p\rangle=\e^{p\cdot V} \left(\frac{\Lambda}{\mu}\right)^{-f(\lambda)\ln(p\cdot V)}\,.
\ee
On the other hand, we can compute $\langle p|W_{V}|p\rangle$  directly in perturbation theory as explained in \cite{Korchemsky:1992xv,Korchemsky89} and from there we expect the leading divergent contribution to be given by,
\be\label{cusped}
\langle p|W_{V}|p\rangle=\e^{p\cdot V}\left(\frac{L}{\epsilon}\right)^{-2\Gamma\ln(p\cdot V)}\,,
\ee
where $L$ and $\epsilon$ are  ultraviolet and infra-red  cut-offs respectively.
By doing the identification $2\Gamma =f(\lambda)$, we can see that (\ref{limitseries}) and (\ref{cusped}) agree, which means that the leading contribution to the divergence of the expectation value of $W_V$ is coming from the operators in the  power expansion with the largest spin.

We will choose the vector $V^ {\mu}$ along a null direction $x^+$ which means we are taking the Wilson line along the given null direction. That will not necessarily happen for finite spin $S$, but  it is certainly the case for infinite spin limit. In that limit, correlators of Wilson lines of the type above should be dominated by the operators with large spin. Hence we will consider correlators of Wilson lines along null directions. Alternatively, as it has been argued in \cite{Alday:2010ku} and can be deduced through $AdS$/CFT considerations, the insertion of an operator ${\cal O}_S$ produces a displacement between fields $\Phi$ along the $x^+$ direction joined by an adjoint Wilson line connecting the two-operators. Moreover, the larger the spin the larger the distance between fields and they become displaced along the null direction as the spin goes to infinity.  

In this section we will follow the lines of \cite{Korchemsky:1992xv} in order to compute the correlator between Wilson lines (\ref{wilsonxp}) along null directions. Let us suppose one of the line operators emits a couple of particles of momentum $p$ and  $p'$ \footnote{The particles $p$ and $p'$ are single particle states of the scalar $\Phi$}, and compute its propagator at one loop. 
Consider the absorption of a gluon with momentum $k$ and gauge potential $A^{\mu}(k)$ emitted by the Wilson line. In the momentum representation, this process contributes to the correlator  the following vertex\footnote{Here we have used the scalar-vector vertex, see appendix  in {\cite{ChengBook} }. }
\be\label{absortion}
g\int\frac{d^4 k}{(2\pi)^4}\frac{(2\,p-k)_{\mu} }{((p-k)^2+i\epsilon)}A^{\mu}(k)\,. 
\ee
In the massless limit or light-like trajectories, this vertex has singularities when $k$ is collinear 
to $p$ and for soft gluons $k\sim0$.  It has been argued in 
\cite{Korchemsky:1992xv} that when the momentum $k$ is collinear to $p$, the components of $A^{\mu}(k)$ 
transverse to $k$ contribute to higher twist and can be neglected in the limit we are considering.
\begin{center}
\includegraphics[scale=0.5]{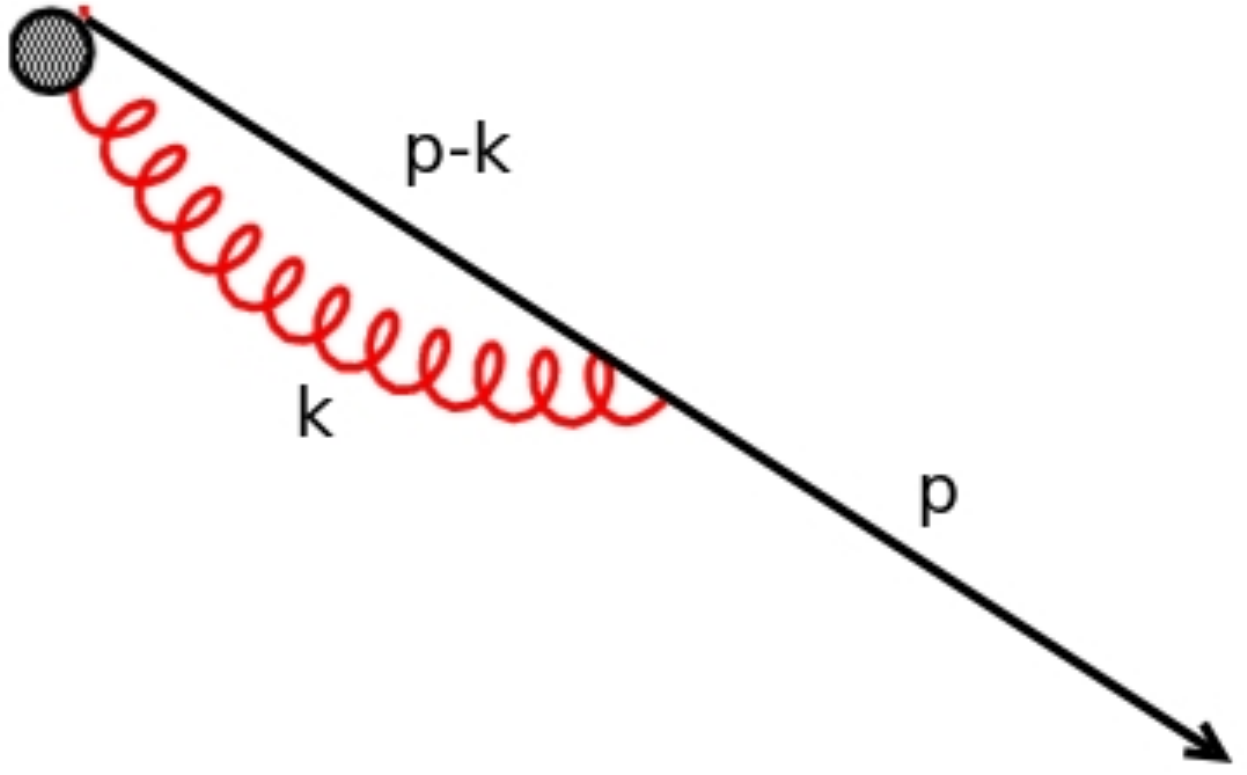}
\end{center}{\small Fig. 1: One loop correction to the particle emission process from a Wilson line}.\newline

Hence only 
longitudinal polarization survives and $A^{\mu}$ becomes pure gauge,
\be\label{gluon}
A^{\mu}(k)=k^{\mu}\left(\frac{y\cdot A(k)}{y\cdot k+i\epsilon}\right)\,,
\ee
Being $y^{\mu}$ a vector along the path followed by the particle. Since the on-shell emitted particle is suppose to be massless, the vector $y^{\mu}$ should be null, i.e, it should be along some $x^-$ direction.  Inserting (\ref{gluon}) into (\ref{absortion}), the absorption vertex is now given by,
\ba
g\int\frac{d^4 k}{(2\pi)^4}\left(\frac{y\cdot A(k)}{y\cdot k+i\epsilon}\right)=g\int_0^{\infty}d\tau\, y_{\mu}A^{\mu}(y\tau)=g\int_0^{\infty}dx^-\cdot A(x^-)\,.
\ea
Summing all contributions at any number of loops of $p'-$collinear gluons, we should finally obtain that the propagator corresponding to the emitted particle in the light-cone limit due to the interaction with gluons turns out to be
\be\label{phase}
{\cal G}(\Delta x^-)=G_{\rm free}(\Delta x^-) {\cal P}{\rm exp}\left(ig\int_{x_1^-}^{x_2^-}dx^-\cdot A(x^-)\right)\, ,
\ee
where $G_{\rm free}$ corresponds to the propagator for the free particle  propagating between points $x_1^-$ and $x_{2}^-$ separated by a light-like distance $\Delta x^-$.
This can be justified in the following way. Since the Wilson line (\ref{wilsonxp}) is dominated by the contributions of local operators at large spin $S$, it could be thought of as a very fast particle propagating in the $x^-$ direction which after a given threshold starts to emit particles and gluons. When the gluons and  particles emitted are collinear, a divergence in the propagator of the emitted particle occurs,  dominating the whole propagator. Since collinear gluons are pure gauge, they cannot change the state of the particle but only  its phase. That pure phase is given by the Wilson line (\ref{phase}) along the light-like direction $x^-$. 

Another way to see this is based on the arguments given in \cite{Alday:2010zy}. The propagator of a scalar particle propagating between points $x_i^-$ and $x_{i+1}^-$ interacting with a gauge field is given by replacing the free scalar propagator by the propagator in the background gluon field $S(x_i^-,x_{i+1}^-;A)$, which satisfies the Green equation
\be
iD^{\mu}D_{\mu}S(x,y;A)=\delta^4(x-y)\,,
\ee
with $D^{\mu}=\partial^{\mu}_{x}-ig[A^{\mu}(x),]$. In the light-cone limit, it is convenient to look for a solution of the above equation of the form (\ref{phase}) times some function of $x$ which goes to one when $x$  tends to $x^{\pm}$.

Concerning to divergences coming from soft glouns, we expect they cancel against virtual gluon corrections as in deep inelastic scatterings in QCD.

In principle, besides the gluon absorption vertex, we should consider scalar absorption vertices also. 
However it has been claimed in \cite{Korchemsky92, Alday:2010zy}, using OPE arguments, that such kind of vertices
 only contribute at higher twist and consequently their contributions are suppressed in the light-like limit.

Finally, putting all the Wilson lines together, i.e, replacing local operators by Wilson lines of the type (\ref{wilsonxp}) and interchanging particle propagators with (\ref{phase}),   the entire correlator becomes a polygonal Wilson loop with light-like edges, which in the very large spin limit we expect to be dominated by the contributions of local operators in the expansion (\ref{wilsonxp}) with large $S$, in a similar way like in the expectation value of a Wilson line with a single cusp {\cite{Kruczenski:2002fb}}. 

\section{Pohlmeyer reduction for strings on $AdS_3$}

In order to set some notation in this section we briefly review the Pohlmeyer reduction of strings on $AdS_3$ proposed in \cite{PohlmeyerNB, DeVegaXC} (see also \cite{Jevicki12, Alday:2009yn} and references therein). We closely follow the notation of \cite{Alday:2009yn}. It is worth mentioning that along this note we will consider only an $AdS_3$ factor of the total $AdS_5$ space. However, we think the results could be generalized to $AdS_5$. 

\vspace{0.2cm}
$AdS_d$ spaces can be written as the following hyperboloid in $R^{2,d-1}$
\be\label{emb}
{\vec{Y}\cdot\vec{Y}=-Y_{-1}^2-Y_{0}^2+Y_1^2+...+Y_{d-1}^2=-1}\, .
\ee
 In terms of these embedding coordinates, the equations of motion for the bosonic string are given by

\be\label{eom}
\partial \bar \partial \vec{Y}-(\partial \vec{Y}\cdot\bar{\partial} \vec{Y})\vec{Y}=0 ~,
\ee
and the Virasoro constraints read
\be
 \partial \vec{Y}\cdot\partial \vec{Y}=\bar \partial \vec{Y}\cdot\bar \partial \vec{Y}=0 \, .
\ee
Let us start defining the reduced fields $\alpha$ and $p$ in $AdS_3$ as,
\ba\label{sinhg}
e^{2 \alpha(z,\bar{z})} &=& \frac{1}{2 } \partial \vec{Y}\cdot\bar \partial \vec{Y}\,, \nn\\
 p =  -\frac{1}{2 } \vec{N}\cdot\partial^2 \vec{Y}~,&& \bar p= \frac{1}{2 }  \vec{N}\cdot{ \bar \partial}^2 \vec{Y}\,, \nn\\
N_a &= &  \frac{ e^{-2 \alpha}}{2 } \epsilon_{abcd}Y^b \partial Y^c \bar \partial Y^d 
\,.
\ea
 From (\ref{sinhg}) and  (\ref{eom}) it can be shown that $p=p(z)$ is a holomorphic function \footnote{For the real solutions considered in this paper, $p(z)$ and $\bar{p}(\bar{z})$ are complex conjugates.
   This condition could in principle be relaxed.}, where we have parametrized the world-sheet in terms of complex variables $z$ and $\bar{z}$. Let us introduce the following basis of four-vectors in $R^{2,2}$
\ba\label{hybridbasis}
\vec{q}_1&= &\vec{Y}~,~~~~~
\vec{q}_2= { e^{-\alpha }  } \bar{\partial} \vec{Y}~,~~~~~
\vec{q}_3={ e^{-\alpha } } {\partial \vec{Y}}~,~~~~~
\vec{q}_4=\vec{N}~,
\ea
which satisfies
\be
\vec{q}_1^{~2}=-1~,~~~~~~~~~~~~\vec{q}_2.\vec{q}_3=2, ~~~~ \vec{q}_4^{~2}=1\,,
\ee
with the remaining $\vec{q}_i.\vec{q}_j=0$. The last property along with the equivalence between $SO(2,2)$ and $SL(2)\times SL(2)$ allows  to write the basis vectors in the following matrix 
representation
\be\label{Wmat}
W= \frac{1}{2 }
\left(\begin{matrix} \vec{q}_1+\vec{q}_4 &\vec{q}_2\\ \vec{q}_3& \vec{q}_1-\vec{q}_4 \end{matrix}\right)\,,
\ee
 W being an $SL(2)\times SL(2)$ element.  The components of this matrix have indices $W_{\alpha \dot{\alpha},a\dot{a}}$. The first two indices denote rows and columns in the above matrix, while the other two are associated with the space-time bispinor representation of each $\vec{q}_i$, i.e,
\be
({q}_i)_{a\dot{a} }=q_{i,\mu}\sigma^{\mu}_{a\dot{a} },\quad \sigma^{\mu}=(1,i\sigma_3,\sigma_1,-\sigma_2)\,.
\ee
The two $SL(2)$ symmetries of the $AdS_3$ target space corresponds to the $SL(2)$ group acting on the index $a$ and the $SL(2)$ that acts on the index $\dot a$. 
The basis vectors (\ref{hybridbasis}) define the $SL(2)$ connections $B^{L,R}$ given by \cite{DeVegaXC}
\ba\label{FGhybrid}
B_z^L=\left(\begin{matrix} \frac{1}{2 } \partial \alpha &&- e^{\alpha} \\
-e^{- \alpha} p(z)  && -\frac{1}{2 } \partial \alpha  \end{matrix}\right),~~~~~ B_{\bar{z}}^L=\left(\begin{matrix}-\frac{1}{2 } \bar\partial\alpha && -e^{-\alpha} {\bar p }(\bar{z})  \cr- e^{ \alpha}    &&\frac{1}{2 } \bar{\partial} \alpha \end{matrix}\right)\\
B_z^R=\left(\begin{matrix}-\frac{1}{2 } \partial \alpha &{e^{-  \alpha} p(z)  }  \\ -{e^{  \alpha}  } &\frac{1}{2 } \partial \alpha\end{matrix}\right),~~~~~
B_{\bar{z}}^R=\left(\begin{matrix} \frac{1}{2 }\bar{\partial} \alpha & -{e^{  \alpha} }  \cr{e^{-  \alpha} {\bar p }(\bar{z}) }  & -\frac{1}{2 } \bar{\partial} \alpha \end{matrix}\right)\, .
\ea
The consistency conditions of the equations (\ref{eom}) imply that these connections are flat
\be
\partial B_{\bar z}^{L}-\bar{\partial} B_z^{L}+[B_z^{L},B_{\bar{z}}^{L}]=0 ~,~~~\partial B_{\bar z}^{R}-\bar{\partial} B_z^{R}+[B_z^{R},B_{\bar{z}}^{R}]=0\, .
\ee
These imply that $p$ is a holomorphic function and that $\alpha$ satisfies the generalized sinh-Gordon equation
\be\label{gensinh}
\partial \bar \partial \alpha(z,\bar{z})-e^{ 2 \alpha(z,\bar{z}) }+|p(z)|^2 e^{-2 \alpha(z,\bar{z})}=0\, .
\ee
We are especially interested in the area of the worldsheet, which is given in terms of the reduced fields by,
\be\label{area}
A =4 \int d^2z  e^{2 \alpha }\,.
\ee
Given a solution of the generalized sinh-Gordon model,
one can find the associated classical string worldsheet. In order to do that, consider the auxiliary linear problem \cite{Alday:2009yn}
\ba\label{LS}
\partial \psi^L_{\alpha}+(B_z^L)_{\alpha}^{~\beta}\psi^L_{\beta}=0,~~~~~
\bar{\partial} \psi^L_{\alpha}+(B_{\bar z}^L)_{\alpha}^{~\beta}\psi^L_{\beta}=0 \, ,\cr
\partial \psi^R_{\dot \alpha }+(B_z^R)_{\dot \alpha}^{~\dot \beta}\psi^R_{\dot \beta } =0,~~~~~\bar{\partial} \psi^R_{\dot \alpha }+(B_z^R)_{\dot \alpha}^{~\dot \beta}\psi^R_{\dot \beta } =0\, .
\ea
We denote by $\psi^L_{\alpha, a}$ , $a=1,2$, and
$\psi^R_{\dot \alpha, \dot a}$, $\dot a =1,2$, the two independent solutions for the left and right linear equations respectively. Since the connections are in $SL(2)$, we can define an $SL(2)$ invariant product and use it to normalize the pair of solutions as
\be\label{normalization}
\langle\psi^L_a,\, \psi^L_b \rangle\equiv  \epsilon^{\beta \alpha } \psi^L_{\alpha,a} \psi^L_{\beta,b}=\epsilon_{a b},~~~~~\langle\psi^R_{\dot{a}},\, \psi^R_{\dot{b}} \rangle\equiv \epsilon^{\dot \beta \dot \alpha } \psi^R_{\dot \alpha, \dot a} \psi^R_{\dot \beta,\dot b}=\epsilon_{\dot a \dot b}\, .
\ee
It is easy to see from (\ref{LS}) that the normalizations (\ref{normalization}) are both constant and then they
can be evaluated at any point.
Now we can reconstruct the space-time worlsheet coordinates from these solutions through the matrix $W_{\alpha \dot \alpha, a \dot a }$, as has been explained in \cite{Alday:2009yn}
\be
\label{Wsol}
{W}_{\alpha \dot \alpha, a \dot a}=\psi^L_{\alpha,a} \psi^R_{\dot \alpha,\dot{a}}\,.
\ee
 This could be justified noticing that each component of $W$ in (\ref{Wmat}) is a null vector in
 $R^{2,2}$, so they can be written as a product of spinors.
The explicit form of the target coordinates is given by the  element $q_1$ in
(\ref{Wmat}), which can be written as

\be
\label{inversemap}  Y_{a \dot a } =
\left(\begin{matrix} Y_{-1}+Y_{2}& Y_1-Y_0\cr Y_1 + Y_0 & Y_{-1}-Y_{2} \end{matrix}\right)_{a,\dot{a}}=
\psi^L_{\alpha,a} \psi^R_{\dot \beta ,\dot{a}} ~,
\ee
and similarly for the other $q_i$.
\section{Large spin limit of the GKP String in reduced fields}
In this section we would like to compute the divergent part of the worldsheet area $A_{\text{div}}$ following an approach similar to the one used in  \cite{Alday:2009yn}. In order to do that we will use the reduced fields just described in the section above in the particular case of a worldsheet which contains asymptotic large spin limit of GKP strings. 

In global coordinates the GKP string at large spin $S$ is given in worldsheet coordinates $(\tau,\,\sigma)$ by  \cite{Roiban:2010fe, {Klebanov02}}
\be\label{global}
t=\kappa \tau,\quad \theta=\omega\tau,\quad\rho=\rho(\sigma)\sim\kappa\sigma,\quad \omega\rightarrow\kappa\sim\frac{1}{\pi}\ln S\\\,.
\ee
In embedding coordinates its looks like
\be
Y_{-1}+iY_0=\e^{\kappa\tau}\cosh\rho(\sigma),\quad Y_1+iY_2=\e^{-\kappa\tau}\sinh\rho(\sigma)\,.
\ee 
As we can see from (\ref{global}), in the limit $S$ going to infinity the parameter $\kappa$ also goes to infinity and therefore it is convenient to make a reparametrization of the cylinder coordinates as,
\be
 \tilde{\tau}=\kappa\tau\,\quad\tilde{\sigma}=\kappa\sigma\,.
\ee
Now the coordinates $\tilde{\tau}\in(-\infty,\infty)$ and $\tilde{\sigma}\in(-\infty,\infty)$.
Inserting the last coordinates in (\ref{sinhg}), we get that the GKP strings are described
 by the following reduced fields in cylindrical coordinates $w=\tilde{\tau}+i\tilde{\sigma}$,
\be
p(w)=-\frac{1}{4},\quad \e^{2\alpha(w,\bar{w})}=\frac{1}{4}\,.
\ee
Going to the plane through the conformal transformation $z=\e^{w}$, we have
\be
p(z)= -\frac{1}{4\,z^2},\quad \e^{2\alpha(z,\bar{z})}=\sqrt{p\bar{p}}\,.
\ee
As we can see, $p(z)$ encodes the information on the vertex positions, which in this case appear at $z=0,\infty$.  
Now  we would like to solve the linear problem (\ref{LS}) associated to the folded string. It is easier to solve it
first in $w$-coordinates, since the reduced fields do not depend on the coordinates there. Doing so we obtain (see \cite{Kazama:2011cp})
\be
 \tilde{\psi}={\cal{A}}\psi=\frac{1}{\sqrt{2} }\e^{\pm\frac{i}{2}(\zeta^{-1}w-\zeta{\bar{w} }) }\left(\begin{matrix}1\\ \pm 1\end{matrix}\right)\,,
\ee
where ${\cal A}=\text{diag}(p^{-1/4}\e^{\alpha/2},p^{1/4}\e^{-\alpha/2})$ and we have introduced a spectral parameter $\zeta$ in order to write the solutions in a compact way. The actual space time solutions are given by  $\zeta=1$ for $\psi^{L}$ and $\zeta=i$ for $\psi^{R}$. Taking $z=r\e^{i\phi}$, on the $z-$plane, the above solutions behave as
\be\label{GKPz}
 \eta^{L\pm}\sim \e^{\pm\phi}\,v_{L\pm},\quad \eta^{R\pm}\sim r^{\pm}\,v_{R\pm}\,,
\ee
where the $v's$ are given by
\be
 v_{L+}=\left(\begin{matrix}-i\\  1\end{matrix}\right),\quad v_{L-}=\left(\begin{matrix}-1\\  i\end{matrix}\right),\quad v_{R+}=\left(\begin{matrix}1\\ 1\end{matrix}\right),\quad v_{R-}=\left(\begin{matrix}-1\\  1\end{matrix}\right)\,.
\ee
Every solution can be written as a linear combination of the above $\eta$ solutions. Hence, we choose the following arbitrary combinations as a basis of solutions
\be\label{basisGKP}
\psi^L_a=c_{l,a}\eta^{L,l},\quad\psi^R_{\dot{a}}=c_{l,\dot{a}}\eta^{R,l},\quad a,\dot{a}=1,2\,. 
\ee
Generically, we are looking for solutions behaving as those above near each singularity produced by the insertion of a vertex operator. Namely, generically we assume the function $p(z)$ has n-singularities at points $z_i,\,i=1,...,n$, describing the insertion of vertex operators in the $z-$plane. More explicitly, we assume the singularities are far enough from each other such that the function $p(z)$ behaves near singularities in the following approximate way
\be\label{pz}
p(z)\sim-\sum_{i=1}^{n} \frac{1 }{4(z-z_i)^2}\, ,
\ee
Hence, near each singular point, the basis of solutions behaves as
\be\label{Nbasis}
\psi^L_{a\,i}\sim c_{l,a\,i}\,\eta^{L,l},\quad\psi^R_{\dot{a}\,i}\sim c_{l,\dot{a}\,i}\,\eta^{R,l},\quad a,\dot{a}=1,2,\quad i=1,...,n-1\,.
\ee
In order to see how the solutions approach each vertex insertion in the generic case, it is convenient to write them in $(z,\bar{z})$ coordinates
\be\label{etageneric}
\eta^{R\,\pm}\sim\prod_{i=1}^n[(z-z_i)(\bar{z}-\bar{z}_i)]^{\pm 1/2},\quad\eta^{L\,\pm}\sim\prod_{i=1}^n\left(\frac{z-z_i}{\bar{z}-\bar{z}_i}\right)^{\pm i/2}\,.
\ee
The target space coordinates can be recovered from the above solutions using (\ref{inversemap}), such that in Poincare  coordinates $(Z, x_{\mu})$ we have
\be\label{poincarecoord}
\frac{1}{Z} =Y_{1\dot{1}},\quad x^{+}=\frac{Y_{1\dot{2}} }{Y_{1\dot{1}}},\quad x^{-}=\frac{Y_{2\dot{1}} }{Y_{1\dot{1}}},\quad x^ {\pm}= x_0\pm x_1\,.
\ee 
As we see from (\ref{etageneric}), some solutions get bigger and some get smaller when we approach the vertex. Therefore the target coordinates 
will be dominated by the big solutions when evaluated near the insertions. Then, from (\ref{inversemap}) and (\ref{Nbasis}), they are schematically approximated near the singularities by
\be\label{bigsolutions}
Y_{a\dot{a} }= c^{L\,\text{big} }_{a\,i}c^{R\,\text{big} }_{\dot{a}\,i}f_{\text{big} }(z-z_i,\bar{z}-\bar{z}_i)\,, 
\ee
where the label 'big' on the coefficients means we keep only those multiplying the $\eta^{R,L}$ which become bigger near the given $z_i$ and $f_{\text{big} }(z-z_i,\bar{z}-\bar{z}_i)$ is the large contribution coming from the $\eta^{R,L}$ which grows 
up the most.
\be\label{Poincare}
 \frac{1}{Z}=c^{L\,\text{big} }_{1\,i}c^{R\,\text{big} }_{\dot{1}\,i}f_{\text{big} }(z-z_i,\bar{z}-\bar{z}_i)\,,\quad
x^{+}=\frac{c^{R\,\text{big} }_{\dot{2}\,i}}{c^{R\,\text{big} }_{\dot{1}\,i}}\,,\quad
x^{-}=\frac{c^{L\,\text{big} }_{{2}\,i}}{c^{L\,\text{big} }_{{1}\,i}}\,.
\ee
\subsection{Divergent area $A_{\text{div}}$}

As we already mentioned, the source of divergences for the area comes from the regions close to the insertions,  which are precisely the regions where the string approaches the boundary of $AdS$. Therefore, it is obvious that putting a worldsheet cutoff near each singularity is completely equivalent to putting a cutoff in target space for the coordinate $Z$ close to the boundary, which at the same time 
corresponds, through $AdS$/CFT,  to putting an ultraviolet cutoff on the energy of the process. In this section we will show how the divergent contribution to the correlator (\ref{1.4}) at strong coupling, $A_{\text{div}}$, is related to the space-time cutoff and how it depends on the dynamics of the string at the boundary. \\

From equations (\ref{GKPz}) and (\ref{Nbasis}) we can see the large spin solutions diverge near
 $\quad\quad\quad\phi=\pm\infty,\, r=\pm\infty$, and therefore, the worldsheet is approaching the boundary at those points. Moreover, as we will see, the worldsheet approaches the boundary on  light-like trajectories. This can be seen from  equation (\ref{Poincare}), realizing that some coefficients $c^{{\rm big} }_{a,i}$ are equal for consecutive $i$, i.e, the points $i$ and $i+1$ have the same, lets say $x^+_i$ coordinate and therefore they are joined by a null line.  This fact will allow us to compute the divergent part of the correlator in the particular kinematic configuration when all the vectors $V^{\mu}$ in (\ref{Twistwo}) are parallel to the $x^+$ direction (see Fig. 2) and are joined by vectors parallel to the $x^-$ direction.
 
The area is given by (\ref{area}), and the divergent part of it comes from the regions near singularities, and as we said, near each singularity $\e^{2\alpha}\sim \sqrt{p\bar{p}}$. We will regularize the area by using a radial cut-off around each singularity $|z-z_i|>\epsilon_i$. Moreover, as we can see from the mapping $\e^w=z,$ the $z-$plane is actually an infinite covering of the complex plane, and the integral over $\phi$ will introduce another source of divergence for the area, and then we should regularize that by putting a cut-off $\Lambda_{\phi}$ in $\phi$.  
Since the leading contribution to the divergent area comes from the regions very near the singularities of $p(z)$, we are going to isolate the contributions of each $z_i$ and approximate $A_{\text{div} }$ as
\be\label{aaprox}
A_{\text{div} }=  4\sum_{i=1}^n\int_{|z-z_i|>\epsilon_i} d^2\,z\,\frac{1}{|z-z_i|^2}\sim -4\sum_{i=1}^n \Lambda_{\phi_i}\ln \epsilon_i\,,
\ee
As we mentioned, when we approach the singular points $z_i$, the worldsheet gets closer to the boundary, and hence the worldsheet cutoff $\epsilon$ should be related to some physical cutoff  $1/Z=1/{\mu},\, \mu<<1$, which corresponds to putting a brane very close to the boundary where the tips of the folded string will end in the limit of infinite spin.  In (\ref{aaprox}) we have left a label $i$ for the cut-off in order to track the corresponding singularity which will be associated to a given space-time coordinate at the boundary. It will become clearer below.

Without loss of generality, let us start considering only three operator insertions, taking $\kappa_1,\,\kappa_3\,>0$ and $\kappa_2\,<0$. As we see from equation (\ref{bigsolutions}), the behaviour of the target coordinates near the boundary is well approximated by the big solutions near each vertex. 
In order to visualize the behavior of the target space coordinates near the insertions, we display the figures 2 and  3. 

\begin{center}
\begin{center}
\includegraphics[scale=0.5]{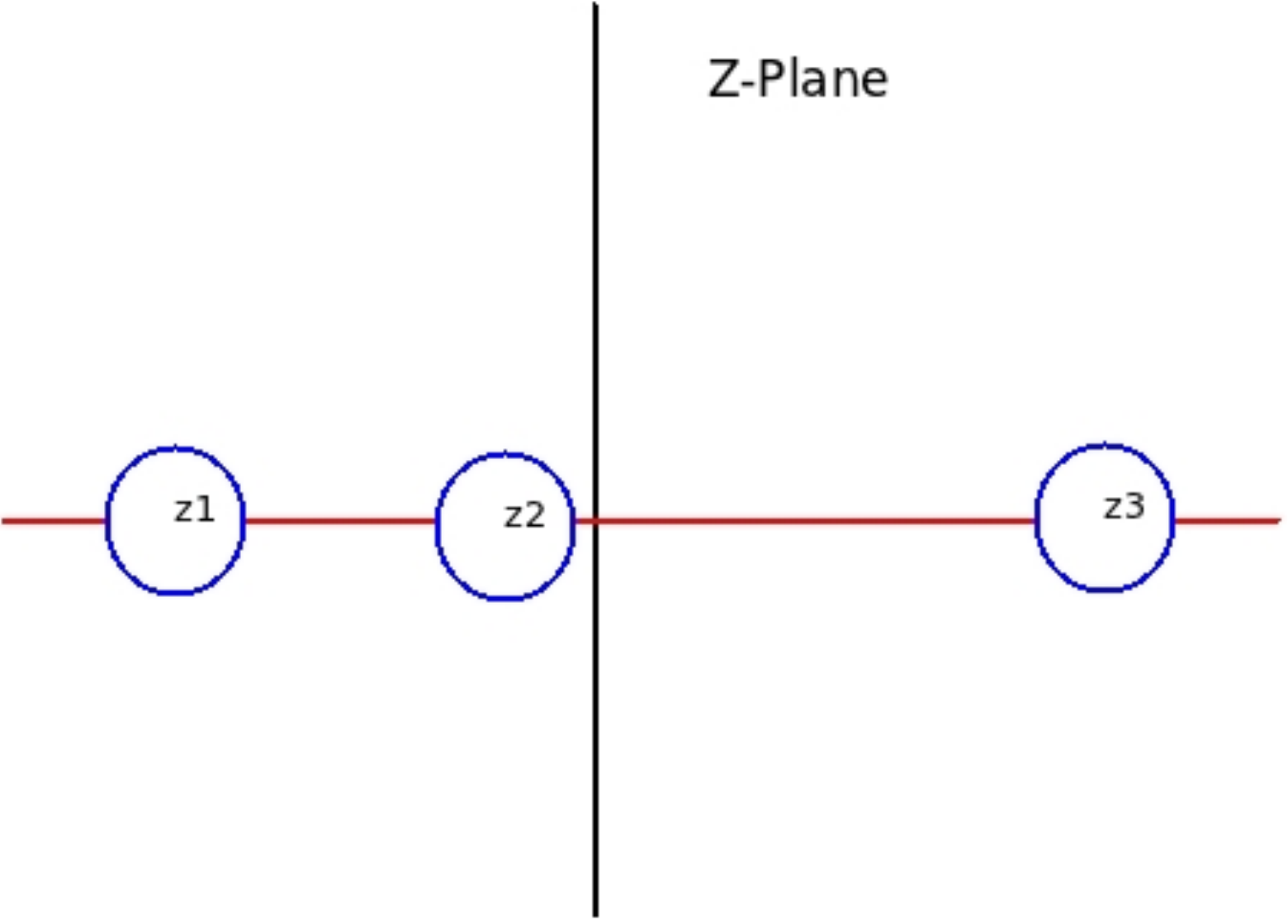}
\end{center}
 fig.2:
$z-$plane with three holes of size $\epsilon$ representing the position of the insertions.
\end{center}
As we will see, red lines and blue lines map to the null directions defining the Wilson loop. Therefore, the points $1,2,...,6$ correspond in  space-time to the cusp of the null Wilson loop. Let us denote the positions of those cusps as $(Z_l,\,x^{\pm}_l)$, and see how we approach the boundary around each cusp. Since $\kappa_1,\,\kappa_3\,>0$ and $\kappa_2\,<0$, we have according to (\ref{bigsolutions}) and (\ref{Poincare}),
\ba\label{cupscoord}
 \frac{1}{Z_1}=c^{L\,+}_{1\,1}c^{R\,-}_{\dot{1}\,1}\left(\frac{z-z_1}{\bar{z}-\bar{z}_1}\right)^{i/2}|z-z_1|^{-1},\quad
x_1^{+}=\frac{c^{R\,-}_{\dot{2}\,1}}{c^{R\,-}_{\dot{1}\,1}}\,,\quad
x_1^{-}=\frac{c^{L\,+}_{{2}\,1}}{c^{L\,+}_{{1}\,1}}\,,\nn\\
\frac{1}{Z_2}=c^{L\,-}_{1\,1}c^{R\,-}_{\dot{1}\,1}\left(\frac{z-z_1}{\bar{z}-\bar{z}_1}\right)^{-i/2}|z-z_1|^{-1},\quad
x_2^{+}=\frac{c^{R\,-}_{\dot{2}\,1}}{c^{R\,-}_{\dot{1}\,1}}\,,\quad
x_2^{-}=\frac{c^{L\,-}_{{2}\,1}}{c^{L\,-}_{{1}\,1}}\,,\nn\\
\frac{1}{Z_3}=c^{L\,+}_{1\,2}c^{R\,+}_{\dot{1}\,2}\left(\frac{z-z_2}{\bar{z}-\bar{z}_2}\right)^{i/2}|z-z_2|^{-1},\quad
x_3^{+}=\frac{c^{R\,+}_{\dot{2}\,2}}{c^{R\,+}_{\dot{1}\,2}}\,,\quad
x_3^{-}=\frac{c^{L\,+}_{{2}\,2}}{c^{L\,+}_{{1}\,2}}\,,\\
\frac{1}{Z_4}=c^{L\,-}_{1\,2}c^{R\,+}_{\dot{1}\,2}\left(\frac{z-z_2}{\bar{z}-\bar{z}_2}\right)^{-i/2}|z-z_2|^{-1},\quad
x_4^{+}=\frac{c^{R\,+}_{\dot{2}\,2}}{c^{R\,+}_{\dot{1}\,2}}\,,\quad
x_4^{-}=\frac{c^{L\,-}_{{2}\,2}}{c^{L\,-}_{{1}\,2}}\,,\nn\\
\frac{1}{Z_5}=c^{L\,+}_{1\,3}c^{R\,-}_{\dot{1}\,3}\left(\frac{z-z_3}{\bar{z}-\bar{z}_3}\right)^{i/2}|z-z_3|^{-1},\quad
x_5^{+}=\frac{c^{R\,-}_{\dot{2}\,3}}{c^{R\,-}_{\dot{1}\,3}}\,,\quad
x_5^{-}=\frac{c^{L\,+}_{{2}\,3}}{c^{L\,+}_{{1}\,3}}\,,\nn\\
\frac{1}{Z_6}=c^{L\,-}_{1\,3}c^{R\,-}_{\dot{1}\,3}\left(\frac{z-z_3}{\bar{z}-\bar{z}_3}\right)^{-i/2}|z-z_3|^{-1},\quad
x_6^{+}=\frac{c^{R\,-}_{\dot{2}\,3}}{c^{R\,-}_{\dot{1}\,3}}\,,\nn\quad
x_6^{-}=\frac{c^{L\,-}_{{2}\,3}}{c^{L\,-}_{{1}\,3}}\nn\,.
\ea

As we  mentioned in section 4, we can see from the equations above that the target positions of points connecting red lines on the worldsheet (see fig.2) live on the same null lines in space-time, i.e $x_1^+=x_2^+,\,x_3^+=x_4^+,\,x_5^+=x_6^+$. 
\begin{center}
\begin{center}
\includegraphics[scale=0.5]{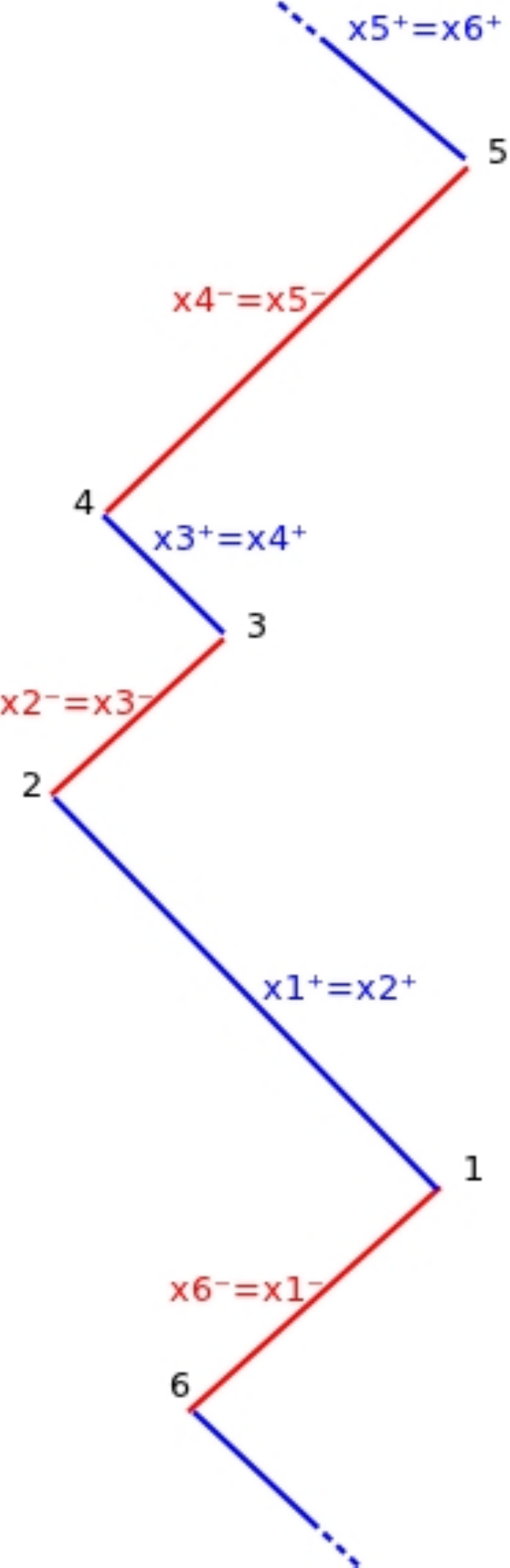}
\end{center}
Fig. 2: {\small Mapping of the boundaries of the half-worlsheet into space-time light-like lines.}
\end{center}
The coordinates displayed in (\ref{cupscoord}) are good approximations to the exact string solution only in the neighborhood of each insertion (think about each insertion as an asymptotic state in target space). Each asymptotic state is connected to each other through the worldsheet, but since we do not know the exact solution, we cannot determine how these states are exactly correlated. One way to relate each insertion to each other is by  using the so-called monodromy matrix.  
As the auxiliary linear problem has two independent solutions, with a basis given by $(\psi_1,\psi_2)$, we expect that as one analytically continues them around a singularity, they get linearly transformed 
as
\be\label{monodromy_act}
\left(\begin{matrix}
\psi_1' \\ \psi_2'
\end{matrix}\right)=S\left(\begin{matrix}
\psi_1 \\ \psi_2
\end{matrix}\right)\,.
\ee
It is worth  mentioning that the monodromy matrix should belong to $SL(2, R),$\footnote{We have written  $\vec{Y}\,\in\,$SO(2,2) as an element  $Y_{a\dot{a}}\sim\psi_a^L\psi_{\dot{a}}^ R$ of SL(2,R)$\times$SL(2,R), or in other words, the eigenvectors of $S$ should live on a representation of SL(2,R).} and therefore it satisfies ${\rm det}|S|=1$. Applying (\ref{monodromy_act}) to (\ref{Nbasis}), we see that we can reinterpret (\ref{monodromy_act}) as
\be
c^{L\alpha}_{a,i}=S^{L\alpha}_{\beta}c^{L\beta}_{a,i}\,,
\ee
with a similar relation for the $c^R$ coefficients and  $\alpha,\,\beta=\pm$.

According to the discussion in section 2, we expect that the leading contribution to the correlator comes from the particles propagating between the operators through nearly null trajectories. Let us start imposing a null trajectory condition to the path connecting  cusp two to  cusp three
\be\label{nullcondition}
x^-_2=\frac{c^{L-}_{2,1}}{c^{L-}_{1,1}}\equiv x^-_3=\frac{c^{L+}_{2,2}}{c^{L+}_{1,2}}=\frac{S^{L+}_{\beta,\,2\rightarrow 1}c^{L\beta}_{2,1}}{S^{L+}_{\beta,\,2\rightarrow 1}c^{L-}_{1,1}}\,,
\ee  
therefore, the $S^{L}-$matrix should satisfy $S^{L+}_{+,\,2\rightarrow 1}=0$, and from the determinant we get $S^{L+}_{-,\,2\rightarrow 1}S^{L-}_{+,\,2\rightarrow 1}=-1$, i.e

\be\label{s21}
S^L_{2\rightarrow 1}=
\left(\begin{matrix}
0 && \pm 1\\ \mp 1 && \gamma^L_{2\rightarrow 1}
\end{matrix}\right)\,,
\ee 
where we have defined $S^{L-}_{-,\,2\rightarrow 1}=\gamma^{L}_{2\rightarrow 1}$. Doing the same for the points connecting cusps 4 and 5 and cusps 6 and 1, we end up with\footnote{These monodromy matrices take the form of the Stokes matrices considered in \cite{Alday:2009yn} by using a different argument. }

\be
S^L_{3\rightarrow 2}=
\left(\begin{matrix}
0 && \pm 1\\ \mp 1 && \gamma^L_{3\rightarrow 2}
\end{matrix}\right)\,,\quad 
S^L_{1\rightarrow 3}=
\left(\begin{matrix}
0 && \pm 1\\ \mp 1 && \gamma^L_{1\rightarrow 3}
\end{matrix}\right)\,.
\ee 
Now we would like to compute the distances between consecutive points, or the separation length of points living on a null line. Let us start considering 
\be
x_1^--x_2^-=\frac{c^{L+}_{2,1}c^{L-}_{1,1}-c^{L+}_{{1},1}c^{L-}_{2,1}}{c^{L+}_{{1},1}c^{L-}_{1,1}} =-\frac{\langle\psi^L_{1,1}\,,\,\psi^L_{2,1}\rangle}{c^{L+}_{{1},1}c^{L-}_{1,1}}\,.
\ee
Recalling we have normalized the basis of solutions in each sector as $\langle\psi^L_{1,i}\,,\,\psi^L_{2,i}\rangle=1$, we get\footnote{We define $x_{i,j}^ {\pm}=x_{i}^ {\pm}-x_{j}^ {\pm}$.}
\be\label{x12}
x_1^--x_2^- =-\frac{1}{c^{L+}_{{1},1}c^{L-}_{1,1}}\equiv x_{12}^-\,.
\ee
On the other hand, we have $x_2^-\equiv x_3^-$, and then
\be\label{x23}
 x_2^+-x_3^+=\frac{c^{R-}_{\dot{2},1}}{c^{R-}_{\dot{1},1}}-\frac{c^{R+}_{\dot{2},2}}{c^{R+}_{\dot{1},2}}\equiv
\frac{a^R_{\dot{2},\dot{1}}}{c^{R-}_{\dot{1},1}c^{R+}_{\dot{1},2}}\,,
\ee
where we have defined
\ba\label{a}
 a^R_{\dot{2},\dot{1}}&=&c^{R-}_{\dot{2},1}({S^{R+}_{+\,2\rightarrow 1}c^{R+}_{\dot{1},1}+S^{R+}_{-\,2\rightarrow 1}c^{R-}_{\dot{1},1}})-c^{R-}_{\dot{1},1}({S^{R+}_{+\,2\rightarrow 1}c^{R+}_{\dot{2},1}+S^{R+}_{-\,2\rightarrow 1}c^{R-}_{\dot{2},1}} )\nn\\&=&\langle\psi^R_{1,1}\,,\,\psi^R_{2,1}\rangle\,S^{R+}_{+\,2\rightarrow 1}= \gamma^{R}_{2\rightarrow 1}\,,
\ea
with $S^{R+}_{+\,2\rightarrow 1}= \gamma^{R}_{2\rightarrow 1}$. The parameters $\gamma^{R,L}_{i+1\rightarrow i}$  parametrize our ignorance on the exact solution and we cannot compute them only using the approximated solutions around the insertions. Notice that in the notation we have used we get $x^-_{12}=x^-_{13}=x^-_{62}=x^-_{63}$ and similar relations for the other cusps (see Fig. 2).\\
Introducing a cutoff $\mu_i$ given by the value of $Z_i$ closest to the boundary and recalling that the worldsheet radial cutoff is defined by $|z-z_i|>\epsilon_i$ we have from (\ref{cupscoord})
\ba\label{mu1}
-\ln\mu_1=\ln(c^{L+}_{{1},1}c^{R-}_{\dot{1},1})-\ln\epsilon_1\,,\\
-\ln\mu_3=\ln(c^{L-}_{{1},2}c^{R-}_{\dot{1},2})-\ln\epsilon_2\label{mu2}\,,
\ea
where we have taken an arbitrary finite $\phi$.
Summing up (\ref{mu1}) and (\ref{mu2}) and using (\ref{x12}), we get
\be\label{logepsilon}
-\ln\epsilon_1\epsilon_3=-\ln\mu_1\mu_3+\ln\left(x_{13}^-x_{13}^+\right)+\ln\gamma^{R}_{2\rightarrow 1}=-\ln\mu_1\mu_3+\ln\left(x_{12}^-x_{23}^+\right)+\ln\gamma^{R}_{2\rightarrow 1}\,.
\ee

In order to disentangle the divergence coming from small radius $\epsilon$ from the one coming from large $\phi$, we re-evaluate (\ref{mu1}) and (\ref{mu2}) at arbitrary but not small  $|z-z_i|$. Then, we have,
\be\label{lambdaphi}
-\Lambda_{\phi_1}-\Lambda_{\phi_3}=-\ln\mu_1\mu_3+\ln\left(x_{13}^-x_{13}^+\right)+\ln\gamma^{R}_{2\rightarrow 1}=-\ln\mu_1\mu_3+\ln\left(x_{12}^-x_{23}^+\right)+\ln\gamma^{R}_{2\rightarrow 1}\,.
\ee
Doing the same for the other lines connecting consecutive cusps, we can see that,
\ba
-\ln\epsilon_3\epsilon_5\sim-\Lambda_{\phi_3}-\Lambda_{\phi_5}&=&-\ln\mu_3\mu_5+\ln\left(x_{34}^-x_{45}^+\right)+\ln\gamma^{R}_{3\rightarrow 2}\nn\\
-\ln\epsilon_5\epsilon_1\sim-\Lambda_{\phi_5}-\Lambda_{\phi_1}&=&-\ln\mu_5\mu_1+\ln\left(x_{56}^-x_{61}^+\right)+\ln\gamma^{R}_{1\rightarrow 3}\,.\nn
\ea
We should take the same cut-off in all directions, i.e. $\mu_i=\mu,\,\epsilon_i=\epsilon,\,\Lambda_{\phi_i}=\Lambda_{\phi}$ and putting this in (\ref{aaprox}) we get,
\ba
&&A_{\text{div} }\sim\nn\\
&&\left[\ln\left(\frac{x_{12}^-x_{23}^+}{\mu^2}\right)+\ln\gamma^{R}_{2\rightarrow 1}\right]^2+\left[\ln\left(\frac{x_{34}^-x_{45}^+}{\mu^ 2}\right)+\ln\gamma^{R}_{3\rightarrow 2}\right]^2+\left[\ln\left(\frac{x_{56}^-x_{61}^+}{\mu^ 2}\right)+\ln\gamma^{R}_{1\rightarrow 3}\right]^2\,.\nn\\
\ea
For the sake of simplicity, let us change the cusp indexing by line indexing in the following way, 
\be\label{boldx}
x_{12}^-\equiv\mathbf{x}_{1}^-,\,~x_{23}^+\equiv\mathbf{x}_2^+,\,~x_{34}^-\equiv\mathbf{x}_3^-,\,~x_{45}^+\equiv\mathbf{x}_4^+,\,~x_{56}^-\equiv\mathbf{x}_5^-,\, ~x_{61}^+\equiv\mathbf{x}_6^+\,. 
\ee
Additionally, we are going to assume the contributions from the monodromy matrices  $\ln\gamma^{R}_{i\rightarrow j}$ are finite and can be neglected with respect to the factors containing $\mu^2$. Doing the above and generalizing for $n-$points or $2n-$cusp (lines) we have.

\be\label{ultimateAdiv}
A_{\text{div} }\sim  \sum_{i=1}^n  \left[\ln[\frac{\mathbf{x}_i^-\mathbf{x}_{i+1}^+}{\mu^2}]\right]^2\,,
\ee
where $n+1=1$. 

The most remarkable outcome of the above computation is that the leading divergent factor $\left[\ln{\frac{1}{\mu^2} }\right]^2$ has the same scaling in terms of the cut-off as the corresponding divergence of a null polygonal Wilson loop \cite{Alday:2009yn}. This type of log square divergences are also usual in the collinear limit of deep inelastic scattering amplitudes in QCD. It is also worth  mentioning that although the vertices are inserted at $n-$points on the worldsheet, they map to $n-$lines on the boundary of $AdS$ (blue lines, fig. 2) very much like space-time points map to lines in twistor space. However, those lines in the boundary do not touch each other, but are connected by other light-like lines (red lines, fig. 2) which come from the region $\phi\to\pm\infty$ on the worldsheet. Based on this, we suggest that an $n-$point function of single trace twist-two large spin operators can be computed from the expectation value of a polygonal Wilson loop when the spins of the operators go to infinity.

Surely the correlator depends on the quantum numbers of the operators, such as the spin and energy. In the limit considered in (\ref{ultimateAdiv}) these quantities are encoded in the length of, lets say, red lines in fig. 2 as well as in the position of the operators. It would be interesting to study the exact mapping from spins and positions of the operators to adjoint Wilson lines.  

\subsection{Contribution from vertex operators}
Currently, a complete understanding of the vertex operator corresponding to a GKP string state is still missing. However, for the purpose of this section we would like to consider the vertex suggested in \cite{bt1}. In Poincar\'e coordinates it is given by,
\be
V_{\Delta_i}(x^+_i,x^-_i)\sim\left[Z(z)+\frac{|x^+(z)x^-(z)-x^+_ix^-_i|^2}{Z(z)}\right]^{-\Delta_i}\,.
\ee 
Where the index $i$ corresponds to the vertex insertion label, i.e, $(x^+_i,x^-_i)$ are the coordinates on the boundary where the vertex is inserted.

As should be, the vertex diverges on the vertex insertion.  
As we have argued along this note, when the spin of the string goes to infinite (so the conformal weight $\Delta$) the string reach the boundary in the insertion points, i.e,
\be
\lim_{z\to z_i}Z(z)=\mu\,,\ee
being $\mu$ the cut-off we have introduced in section above.

Near the $i-$th vertex insertion $(x^+_i,x^-_i)$, the particular contribution to the correlator is them given by (see eq. (\ref{1.4})),
\be\label{vertexdiv}
\lim_{z\to z_i}\left[-\Delta\ln |x^+(z)x^-(z)-x^+_ix^-_i|^2+\Delta\ln Z(z)\right]=-\Delta\ln\tilde{\epsilon}+\Delta\ln\mu\,,
\ee
where we have defined $|x^+(z_i)x^-(z_i)-x^+_ix^-_i|^2=\tilde\epsilon$, which measures the distance between the worldsheet evaluated on the 'cut-off brane' and the insertion point, and it should be the same order of the cut-off $\mu$.

From (\ref{vertexdiv}) we can see that near the vertex insertion,  the vertex operator is finite and therefore, does not contribute to the divergent factor of the correlator.

\section{Conclusions} 

In this paper we have considered a light-like limit of correlation functions involving operators of high spin. We have 
argued that in the large spin limit, correlators should be related to expectation values of Wilson loops along paths lying on the light cone. At strong 't Hooft coupling we performed a semi-classical computation of the divergent part of the correlator and found that it scales with the cut-off in the same way the expectation value of a null Wilson loop does. 

In order to confirm the correspondence between Wilson loops and high spin correlation functions, at least at semi-classical level, we should find an exact classical solution of a string in $AdS$ which contains GKP classical states and has
 the topology of a sphere with n-vertex insertions. As far as we know, that solution is unknown. However,
the correlator could be computed semi-classically by using integrability,
 as has been done at strong coupling for the finite part in \cite{Kazama:2011cp} and at weak coupling in \cite{Escobedo:2010xs}. The next interesting step could be to use the results in \cite{Kazama:2011cp} in order to extract the finite part of the Wilson loops from the correlators.

\section*{Acknowledgments }

I am grateful to Martin Kruczenski and Juan Maldacena for very useful discussions and interesting comments and suggestions.  I also thank Carmen Nu\~nez who encouraged me to publish this note and for carefully read the manuscript.  I would also like to thank Oscar Bedoya and  Humberto Gomez for reading the first version of this note. I also like to thank to the Institute for Advanced Studies at Princeton for hospitality during the early stages of this work. 

This work was supported by CONICET PIP 112-200801-00507, MINCYT
(Ministerio de Ciencia, Tecnolog\'ia Innovaci\'on Productiva de Argentina), University of Buenos Aires UBACyT X161.

\appendix


\end{document}